\documentclass[10pt,conference]{IEEEtran}
\IEEEoverridecommandlockouts

\usepackage{cite}
\usepackage[T1]{fontenc}
\usepackage{amsmath,amssymb,amsfonts}
\usepackage{algorithmic}
\usepackage{graphicx}
\usepackage{textcomp}
\usepackage{xcolor}
\def\BibTeX{{\rm B\kern-.05em{\sc i\kern-.025em b}\kern-.08em
    T\kern-.1667em\lower.7ex\hbox{E}\kern-.125emX}}

\usepackage[caption=false,font=footnotesize]{subfig}
\usepackage[bookmarks=false]{hyperref}
\usepackage{tikz}
\usetikzlibrary{shapes,arrows}
\usetikzlibrary{positioning,calc}
\usepackage{pgfplots}
\pgfplotsset{compat=1.18}
\usepackage{booktabs}

\begin{document}

\title{Quota-Based Traffic Classification for Bounded Jitter in Multi-CQF under Redundancy Mechanisms}

\author{
  \IEEEauthorblockN{Mohamed Elattar\IEEEauthorrefmark{1}\IEEEauthorrefmark{2}, Zhe Lou\IEEEauthorrefmark{1}, Georg Carle\IEEEauthorrefmark{2}}
  \IEEEauthorblockA{\IEEEauthorrefmark{1}Huawei Heisenberg Research Center, Munich, Germany}
  \IEEEauthorblockA{\IEEEauthorrefmark{2} Technical University of Munich, Munich, Germany}
    \IEEEauthorblockA{\IEEEauthorrefmark{1}\{mohamed.osama.elattar, zhe.lou\}@huawei.com, 
    \IEEEauthorrefmark{2}m.elattar@tum.de,
    \IEEEauthorrefmark{2}carle@net.in.tum.de } 

\thanks{This work has been submitted to the IEEE for possible publication. Copyright may be transferred without notice, after which this version may no longer be accessible.}
}

\maketitle
\begin{abstract}

The demand for deterministic and reliable communication is rapidly increasing in industrial networks. Time-Sensitive Networking (TSN) addresses these requirements through deterministic scheduling and redundancy mechanisms such as Multiple Cyclic Queuing and Forwarding (Multi-CQF) and Frame Replication and Elimination for Reliability (FRER). Multi-CQF guarantees bounded end-to-end (E2E) latency and jitter, while FRER provides seamless fault tolerance through redundant forwarding. With static traffic classification, redundant paths with different hop counts can exhibit different E2E latency ranges. This can amplify jitter when a failure causes packets to be delivered over a longer redundant path. To address this, we propose a quota-based traffic classification mechanism that adapts class assignments hop-by-hop using a common E2E latency budget derived from the primary path, thereby mitigating jitter amplification. Simulation results show that the proposed mechanism reduces jitter amplification by up to 70\% and latency degradation by up to 41\% compared to static classification.

\end{abstract}


\section{Introduction} 
\label{Introduction}

The demand for deterministic and reliable communication is rapidly increasing across industrial automation, vehicular networks, and other cyber-physical systems. These systems require not only strict end-to-end (E2E) latency and jitter bounds, but also continued operation in the presence of failures. IEEE 802.1 Time-Sensitive Networking (TSN)~\cite{8021Q2022} addresses these requirements and introduces mechanisms that provide quality-of-service (QoS) guarantees for latency-sensitive applications. These mechanisms comprise traffic scheduling, which ensures bounded latency and jitter, and reliability mechanisms, which ensure seamless fault tolerance in the presence of failures.

Among deterministic schedulers, Multiple Cyclic Queuing and Forwarding (Multi-CQF)~\cite{multi-CQF} provides bounded E2E latency and jitter across multiple traffic classes with heterogeneous latency and bandwidth requirements, making it a promising solution for large-scale deterministic networks. Reliability is achieved through Frame Replication and Elimination for Reliability (FRER) \cite{frer}, which replicates packets over multiple paths and eliminates duplicates at the destination. 

FRER is frequently combined with deterministic schedulers to jointly provide QoS guarantees and seamless recovery. Prior work has focused on analyzing the behavior of FRER \cite{9488750,3171614,9838905}, as well as its worst-case delay bounds when combined with the Asynchronous Traffic Shaper (ATS) \cite{FRER_bounds} and with classical CQF \cite{Multi-path_CQF}. However, the interaction between FRER and Multi-CQF in heterogeneous network topologies remains insufficiently explored. As we show in this paper, combining them degrades the original E2E latency and jitter guarantees of Multi-CQF. This degradation stems from the static traffic classification, which assigns each flow to a fixed traffic class along all paths. When packet replicas traverse redundant paths with unequal lengths, they accumulate different E2E latencies. In failure scenarios, this latency imbalance leads to jitter amplification and abrupt latency shifts at the destination.

To address this limitation, we propose a quota-based traffic classification mechanism for Multi-CQF under FRER. The proposed approach dynamically adapts traffic-class assignments on a hop-by-hop basis using an E2E latency reference derived from the primary path. By regulating the residual latency budget along redundant paths, the mechanism reduces latency differences across routes, thereby mitigating jitter amplification.

The remainder of this paper is organized as follows. Section~II reviews the background, Section~III presents the system model, Section~IV introduces the proposed mechanism, Section~V evaluates its performance, and Section~VI concludes the paper.

\section{Background} 
\label{Related Work Section}

This section first reviews scheduling mechanisms for bounded latency, followed by redundancy mechanisms for fault tolerance.

\subsection{Scheduling Mechanisms}

TSN provides various mechanisms for traffic scheduling and shaping. The Time-Aware Shaper (TAS) achieves strict timing guarantees through precise time synchronization, while the ATS  avoids synchronization but requires per-flow state tracking, introducing overhead in dynamic networks with frequent flow arrivals and departures.

Cyclic Queuing and Forwarding (CQF) instead bounds E2E latency using only cycle time and hop count, independent of traffic dynamics. Multi-CQF \cite{multi-CQF} generalizes CQF to multiple traffic classes with distinct cycle times, enabling heterogeneous latency and bandwidth requirements while preserving deterministic guarantees. Misaligned cycle configurations across classes may, however, violate latency bounds. Hybrid Scheduling with Strict Priority and Packet Urgency (HSSU)~\cite{HSSU} addresses this limitation through urgency-aware scheduling, allowing more flexible cycle configurations.

\subsection{Reliability Mechanisms}

Retransmission-based recovery can introduce variable delays \cite{RSTP}\cite{SPB}. Proactive redundancy mechanisms are therefore preferred for time-critical systems. Parallel Redundancy Protocol (PRP) \cite{PRP} and High-availability Seamless Redundancy (HSR) provide zero recovery time through frame replication over two separate networks and a ring topology, respectively.

FRER supports seamless redundancy in arbitrary network topologies by replicating frames over multiple paths and eliminating duplicates using sequence numbers. Although FRER enhances reliability, forwarding replicas over paths with unequal delays introduces out-of-order delivery and jitter amplification. When combined with a deterministic scheduler, this additional jitter degrades the scheduler's original QoS guarantees. This effect is analyzed in detail in Section \ref{SOTA Section}.

\definecolor{BlueOLL}{HTML}{5B9BD5}
\definecolor{BlueOLM}{HTML}{2E75B6}
\definecolor{BlueOLD}{HTML}{1F4E79}

\definecolor{BlueFillL}{HTML}{DEEBF7}
\definecolor{BlueFillD}{HTML}{BDD7EE}

\definecolor{GreenOLL}{HTML}{70AD47}
\definecolor{GreenOLM}{HTML}{548235}
\definecolor{GreenOLD}{HTML}{385723}

\definecolor{GreenFillL}{HTML}{E2F0D9}
\definecolor{GreenFillD}{HTML}{C5E0B4}

\definecolor{RedOL}{HTML}{C0392B} 
\definecolor{RedFill}{HTML}{F5B7B1}

\definecolor{CollisionOL}{HTML}{16A085} 
\definecolor{CollisionFill}{HTML}{A3E4D7}

\definecolor{WhiteOLL}{HTML}{BFBFBF}
\definecolor{WhiteOLM}{HTML}{A6A6A6}
\definecolor{WhiteOLD}{HTML}{7F7F7F}

\definecolor{WhiteFillL}{HTML}{F2F2F2}
\definecolor{WhiteFillD}{HTML}{D9D9D9}

\definecolor{GrayOLL}{HTML}{A5A5A5}
\definecolor{GrayOLM}{HTML}{7C7C7C}
\definecolor{GrayOLD}{HTML}{525252}

\definecolor{GrayFillL}{HTML}{EDEDED}
\definecolor{GrayFillD}{HTML}{DBDBDB}

\definecolor{YellowOLL}{HTML}{FFC000}
\definecolor{YellowOLM}{HTML}{BF9000}
\definecolor{YellowOLD}{HTML}{7F6000}

\definecolor{YellowFillL}{HTML}{FFF2CC}
\definecolor{YellowFillD}{HTML}{FFE699}

\definecolor{OrangeOLL}{HTML}{ED7D31}
\definecolor{OrangeOLM}{HTML}{C55A11}
\definecolor{OrangeOLD}{HTML}{843C0C}

\definecolor{OrangeFillL}{HTML}{FBE5D6}
\definecolor{OrangeFillD}{HTML}{F8CBAD}

\newcommand{\SinglePathShortestPath}[0]{
\begin{tikzpicture}[
    node/.style={rectangle, draw, minimum width=0.5cm, minimum height=0.2cm},
    device/.style={circle, draw, minimum size=0.5cm},
    layer/.style={rectangle, draw, minimum width=0.4cm, minimum height=0.2cm},
    every path/.style={draw}
  ]

\node[device] (S) {\tiny S};
\node[node, right=1cm of S] (N1) {\tiny 1}; 
\node[node, right=1.76cm of N1] (N2) {\tiny 2}; 
\node[node, right=1.76cm of N2] (N3) {\tiny 3};
\node[device, right=1cm of N3] (D) {\tiny D};

\path (S) -- (N1);
\path (N1) -- (N2);
\path (N2) -- (N3);
\path (N3) -- (D);

\foreach \n in {N1, N3} {
    \node[layer, fill=white, above=0.125cm of \n] (\n L1) {\tiny $T_1 = 10\mu$s};
    \node[layer, fill=white,  above=0cm of \n L1] (\n L2) {\tiny $T_2 = 20\mu$s};
    \node[layer, fill=GreenFillL, above=0cm of \n L2] (\n L3) {\tiny $T_3 = 30\mu$s};
}

\foreach \n in {N2} {
    \node[layer, fill=white, above=0.125cm of \n] (\n L1) {\tiny $T_1 = 10\mu$s};
    \node[layer, fill=white,  above=0cm of \n L1] (\n L2) {\tiny $T_2 = 20\mu$s};
    \node[layer, fill=GreenFillL, above=0cm of \n L2] (\n L3) {\tiny $T_3 = 30\mu$s};
}

\draw[->, thick, shorten >=0.15cm, shorten <=0.15cm] 
($(S)!0!(N1)!3mm!25:(N1)$) -- 
($(N1)!0!(S)!3mm!-25:(S)$)
node[midway, sloped, below, yshift=-0.75mm] {\tiny $P_f = 3$};

\draw[->, thick, shorten >=0.15cm, shorten <=0.15cm] 
    ($(N3)!0!(D)!3mm!25:(D)$) -- 
($(D)!0!(N3)!3mm!-25:(N3)$)
node[midway, sloped, below, yshift=-0.75mm] {\tiny $P_f = 3$};

\draw[->, thick, shorten >=0.15cm, shorten <=0.15cm, color=GreenOLM] 
($(N1)!0!(N2)!3mm!25:(N2)$) -- 
($(N2)!0!(N1)!3mm!-25:(N1)$)
node[midway, sloped, below, yshift=-0.75mm, text=GreenOLM] {\tiny $P_f = 3$};

\draw[->, thick, shorten >=0.15cm, shorten <=0.15cm, color=GreenOLM] 
($(N2)!0!(N3)!3mm!25:(N3)$) -- 
($(N3)!0!(N2)!3mm!-25:(N2)$)
node[midway, below, yshift=-0.75mm, text=GreenOLM] {\tiny $P_f = 3$};

\end{tikzpicture}
}

\newcommand{\MultiPathRPS}[0]{
\begin{tikzpicture}[
    node/.style={rectangle, draw, minimum width=0.5cm, minimum height=0.2cm},
    device/.style={circle, draw, minimum size=0.5cm},
    layer/.style={rectangle, draw, minimum width=0.4cm, minimum height=0.2cm},
    every path/.style={draw}
  ]

\node[device] (S) {\tiny S};
\node[node, right=1cm of S] (N1) {\tiny 1}; 
\node[node, right=1.76cm of N1] (N2) {\tiny 2}; 
\node[node, right=1.76cm of N2] (N3) {\tiny 3};
\node[device, right=1cm of N3] (D) {\tiny D};

\node[node, below=1cm of $(N1)!0.25!(N3)$] (N4) {\tiny 4};
\node[node, below=1cm of $(N1)!0.75!(N3)$] (N5) {\tiny 5};

\node[node, above=1cm of $(N1)!0.15!(N3)$] (N6) {\tiny 6};
\node[node, above=1cm of $(N1)!0.5!(N3)$] (N7) {\tiny 7};
\node[node, above=1cm of $(N1)!0.85!(N3)$] (N8) {\tiny 8};

\path (S) -- (N1);
\path (N1) -- (N2);
\path (N2) -- (N3);
\path (N1) -- (N4);
\path (N4) -- (N5);
\path (N5) -- (N3);
\path (N1) -- (N6);
\path (N6) -- (N7);
\path (N7) -- (N8);
\path (N8) -- (N3);
\path (N3) -- (D);

\foreach \n in {N1} {
    \node[layer, fill=white, above=0.225cm of \n, xshift=-0.75cm] (\n L1) {\tiny $T_1 = 10\mu$s};
    \node[layer, fill=white,  above=0cm of \n L1] (\n L2) {\tiny $T_2 = 20\mu$s};
    \node[layer, fill=OrangeFillL, above=0cm of \n L2] (\n L3) {\tiny $T_3 = 30\mu$s};
}
\foreach \n in {N3} {
    \node[layer, fill=white, above=0.225cm of \n, xshift=0.75cm] (\n L1) {\tiny $T_1 = 10\mu$s};
    \node[layer, fill=white,  above=0cm of \n L1] (\n L2) {\tiny $T_2 = 20\mu$s};
    \node[layer, fill=OrangeFillL, above=0cm of \n L2] (\n L3) {\tiny $T_3 = 30\mu$s};
}

\foreach \n in {N6, N7, N8} {
    \node[layer, fill=white, above=0.125cm of \n] (\n L1) {\tiny $T_1 = 10\mu$s};
    \node[layer, fill=white,  above=0cm of \n L1] (\n L2) {\tiny $T_2 = 20\mu$s};
    \node[layer, fill=OrangeFillL, above=0cm of \n L2] (\n L3) {\tiny $T_3 = 30\mu$s};
}

\foreach \n in {N1} {
    \node[layer, fill=BlueFillL, below=0.225cm of \n, xshift=-0.6cm] (\n L1) {\tiny $T_3 = 30\mu$s};
    \node[layer, fill=white, below=0cm of \n L1] (\n L2) {\tiny $T_2 = 20\mu$s};
    \node[layer, fill=white, below=0cm of \n L2] (\n L3) {\tiny $T_1 = 10\mu$s};
}

\foreach \n in {N3} {
    \node[layer, fill=BlueFillL, below=0.225cm of \n, xshift=0.6cm] (\n L1) {\tiny $T_3 = 30\mu$s};
    \node[layer, fill=white, below=0cm of \n L1] (\n L2) {\tiny $T_2 = 20\mu$s};
    \node[layer, fill=white, below=0cm of \n L2] (\n L3) {\tiny $T_1 = 10\mu$s};
}

\foreach \n in {N4, N5} {
    \node[layer, fill=BlueFillL, below=0.125cm of \n] (\n L1) {\tiny $T_3 = 30\mu$s};
    \node[layer, below=0cm of \n L1] (\n L2) {\tiny $T_2 = 20\mu$s};
    \node[layer, below=0cm of \n L2] (\n L3) {\tiny $T_1 = 10\mu$s};
}

\draw[->, thick, shorten >=0.15cm, shorten <=0.15cm] 
($(S)!0!(N1)!3mm!25:(N1)$) -- 
($(N1)!0!(S)!3mm!-25:(S)$)
node[midway, sloped, below, yshift=-0.75mm] {\tiny $P_f = 3$};

\draw[->, thick, shorten >=0.15cm, shorten <=0.15cm] 
    ($(N3)!0!(D)!3mm!25:(D)$) -- 
($(D)!0!(N3)!3mm!-25:(N3)$)
node[midway, sloped, below, yshift=-0.75mm] {\tiny $P_f = 3$};

\draw[->, thick, shorten >=0.15cm, shorten <=0.15cm, color=GreenOLM] 
($(N1)!0!(N2)!3mm!25:(N2)$) -- 
($(N2)!0!(N1)!3mm!-25:(N1)$)
node[midway, sloped, below, yshift=-0.75mm, , color=GreenOLM] {\tiny $P_f = 3$};

\draw[->, thick, shorten >=0.15cm, shorten <=0.15cm, color=GreenOLM] 
($(N2)!0!(N3)!3mm!25:(N3)$) -- 
($(N3)!0!(N2)!3mm!-25:(N2)$)
node[midway, below, yshift=-0.75mm, text=GreenOLM] {\tiny $P_f = 3$};

\draw[->, thick, shorten >=0.15cm, shorten <=0.15cm, color=BlueOLM] 
($(N1)!0!(N4)!3mm!25:(N4)$) -- 
($(N4)!0!(N1)!3mm!-25:(N1)$)
node[midway, sloped, below, yshift=-0.75mm] {\tiny $P_f = 3$};

\draw[->, thick, shorten >=0.15cm, shorten <=0.15cm, color=BlueOLM] 
($(N4)!0!(N5)!3mm!25:(N5)$) -- 
($(N5)!0!(N4)!3mm!-25:(N4)$)
node[midway, sloped, below, yshift=-0.75mm] {\tiny $P_f = 3$};

\draw[->, thick, shorten >=0.15cm, shorten <=0.15cm, color=BlueOLM] 
($(N5)!0!(N3)!3mm!25:(N3)$) -- 
($(N3)!0!(N3)!3mm!-25:(N5)$)
node[midway, sloped, below, yshift=-0.75mm] {\tiny $P_f = 3$};

\draw[->, thick, shorten >=0.15cm, shorten <=0.15cm, color=OrangeOLM] 
($(N1)!0!(N6)!3mm!25:(N6)$) -- 
($(N6)!0!(N1)!3mm!-25:(N1)$)
node[midway, sloped, below, yshift=-0.75mm] {\tiny $P_f = 3$};

\draw[->, thick, shorten >=0.15cm, shorten <=0.15cm, color=OrangeOLM] 
($(N8)!0!(N3)!3mm!25:(N3)$) -- 
($(N3)!0!(N8)!3mm!-25:(N8)$)
node[midway, sloped, below, yshift=-0.75mm] {\tiny $P_f = 3$};

\draw[->, thick, shorten >=0.15cm, shorten <=0.15cm, color=OrangeOLM] 
($(N6)!0!(N7)!3mm!25:(N7)$) -- 
($(N7)!0!(N6)!3mm!-25:(N6)$)
node[midway, sloped, below, yshift=-0.75mm] {\tiny $P_f = 3$};

\draw[->, thick, shorten >=0.15cm, shorten <=0.15cm, color=OrangeOLM] 
($(N7)!0!(N8)!3mm!25:(N8)$) -- 
($(N8)!0!(N7)!3mm!-25:(N7)$)
node[midway, sloped, below, yshift=-0.75mm] {\tiny $P_f = 3$};

\end{tikzpicture}
}

\newcommand{\MultiPathQuota}[0]{
\begin{tikzpicture}[
    node/.style={rectangle, draw, minimum width=0.5cm, minimum height=0.2cm},
    device/.style={circle, draw, minimum size=0.5cm},
    layer/.style={rectangle, draw, minimum width=0.4cm, minimum height=0.2cm},
    every path/.style={draw}
  ]

\node[device] (S) {\tiny S};
\node[node, right=1cm of S] (N1) {\tiny 1}; 
\node[node, right=1.76cm of N1] (N2) {\tiny 2}; 
\node[node, right=1.76cm of N2] (N3) {\tiny 3};
\node[device, right=1cm of N3] (D) {\tiny D};

\node[node, below=1cm of $(N1)!0.25!(N3)$] (N4) {\tiny 4};
\node[node, below=1cm of $(N1)!0.75!(N3)$] (N5) {\tiny 5};

\node[node, above=1cm of $(N1)!0.15!(N3)$] (N6) {\tiny 6};
\node[node, above=1cm of $(N1)!0.5!(N3)$] (N7) {\tiny 7};
\node[node, above=1cm of $(N1)!0.85!(N3)$] (N8) {\tiny 8};

\path (S) -- (N1);
\path (N1) -- (N2);
\path (N2) -- (N3);
\path (N1) -- (N4);
\path (N4) -- (N5);
\path (N5) -- (N3);
\path (N1) -- (N6);
\path (N6) -- (N7);
\path (N7) -- (N8);
\path (N8) -- (N3);
\path (N3) -- (D);

\foreach \n in {N1} {
    \node[layer, fill=OrangeFillL, above=0.225cm of \n, xshift=-0.75cm] (\n L1) {\tiny $T_1 = 10\mu$s};
    \node[layer, fill=white,  above=0cm of \n L1] (\n L2) {\tiny $T_2 = 20\mu$s};
    \node[layer, fill=white, above=0cm of \n L2] (\n L3) {\tiny $T_3 = 30\mu$s};
}
\foreach \n in {N3} {
    \node[layer, fill=white, above=0.225cm of \n, xshift=0.75cm] (\n L1) {\tiny $T_1 = 10\mu$s};
    \node[layer, fill=white,  above=0cm of \n L1] (\n L2) {\tiny $T_2 = 20\mu$s};
    \node[layer, fill=OrangeFillL, above=0cm of \n L2] (\n L3) {\tiny $T_3 = 30\mu$s};
}

\foreach \n in {N6, N7} {
    \node[layer, fill=white, above=0.125cm of \n] (\n L1) {\tiny $T_1 = 10\mu$s};
    \node[layer, fill=OrangeFillL,  above=0cm of \n L1] (\n L2) {\tiny $T_2 = 20\mu$s};
    \node[layer, fill=white, above=0cm of \n L2] (\n L3) {\tiny $T_3 = 30\mu$s};
}

\foreach \n in {N8} {
    \node[layer, fill=white, above=0.125cm of \n] (\n L1) {\tiny $T_1 = 10\mu$s};
    \node[layer, fill=white,  above=0cm of \n L1] (\n L2) {\tiny $T_2 = 20\mu$s};
    \node[layer, fill=OrangeFillL, above=0cm of \n L2] (\n L3) {\tiny $T_3 = 30\mu$s};
}

\foreach \n in {N1} {
    \node[layer, fill=white, below=0.225cm of \n, xshift=-0.6cm] (\n L1) {\tiny $T_3 = 30\mu$s};
    \node[layer, fill=BlueFillL, below=0cm of \n L1] (\n L2) {\tiny $T_2 = 20\mu$s};
    \node[layer, fill=white, below=0cm of \n L2] (\n L3) {\tiny $T_1 = 10\mu$s};
}

\foreach \n in {N3} {
    \node[layer, fill=BlueFillL, below=0.225cm of \n, xshift=0.6cm] (\n L1) {\tiny $T_3 = 30\mu$s};
    \node[layer, fill=white, below=0cm of \n L1] (\n L2) {\tiny $T_2 = 20\mu$s};
    \node[layer, fill=white, below=0cm of \n L2] (\n L3) {\tiny $T_1 = 10\mu$s};
}

\foreach \n in {N4} {
    \node[layer, below=0.125cm of \n] (\n L1) {\tiny $T_3 = 30\mu$s};
    \node[layer, fill=BlueFillL, below=0cm of \n L1] (\n L2) {\tiny $T_2 = 20\mu$s};
    \node[layer, below=0cm of \n L2] (\n L3) {\tiny $T_1 = 10\mu$s};
}

\foreach \n in {N5} {
    \node[layer, fill=BlueFillL, below=0.125cm of \n] (\n L1) {\tiny $T_3 = 30\mu$s};
    \node[layer, below=0cm of \n L1] (\n L2) {\tiny $T_2 = 20\mu$s};
    \node[layer, below=0cm of \n L2] (\n L3) {\tiny $T_1 = 10\mu$s};
}

\draw[->, thick, shorten >=0.15cm, shorten <=0.15cm] 
($(S)!0!(N1)!3mm!25:(N1)$) -- 
($(N1)!0!(S)!3mm!-25:(S)$)
node[midway, sloped, below, yshift=-0.75mm] {\tiny $P_f = 3$};

\draw[->, thick, shorten >=0.15cm, shorten <=0.15cm] 
    ($(N3)!0!(D)!3mm!25:(D)$) -- 
($(D)!0!(N3)!3mm!-25:(N3)$)
node[midway, sloped, below, yshift=-0.75mm] {\tiny $P_f = 3$};

\draw[->, thick, shorten >=0.15cm, shorten <=0.15cm, color=GreenOLM] 
($(N1)!0!(N2)!3mm!25:(N2)$) -- 
($(N2)!0!(N1)!3mm!-25:(N1)$)
node[midway, sloped, below, yshift=-0.75mm, , color=GreenOLM] {\tiny $\delta_f^1 = 6$};

\draw[->, thick, shorten >=0.15cm, shorten <=0.15cm, color=GreenOLM] 
($(N2)!0!(N3)!3mm!25:(N3)$) -- 
($(N3)!0!(N2)!3mm!-25:(N2)$)
node[midway, below, yshift=-0.75mm, text=GreenOLM] {\tiny $\delta_f^1 = 3$};

\draw[->, thick, shorten >=0.15cm, shorten <=0.15cm, color=BlueOLM] 
($(N1)!0!(N4)!3mm!25:(N4)$) -- 
($(N4)!0!(N1)!3mm!-25:(N1)$)
node[midway, sloped, below, yshift=-0.75mm] {\tiny $\delta_f^2 = 8$};

\draw[->, thick, shorten >=0.15cm, shorten <=0.15cm, color=BlueOLM] 
($(N4)!0!(N5)!3mm!25:(N5)$) -- 
($(N5)!0!(N4)!3mm!-25:(N4)$)
node[midway, sloped, below, yshift=-0.75mm] {\tiny $\delta_f^2 = 6$};

\draw[->, thick, shorten >=0.15cm, shorten <=0.15cm, color=BlueOLM] 
($(N5)!0!(N3)!3mm!25:(N3)$) -- 
($(N3)!0!(N3)!3mm!-25:(N5)$)
node[midway, sloped, below, yshift=-0.75mm] {\tiny $\delta_f^2 = 3$};

\draw[->, thick, shorten >=0.15cm, shorten <=0.15cm, color=OrangeOLM] 
($(N1)!0!(N6)!3mm!25:(N6)$) -- 
($(N6)!0!(N1)!3mm!-25:(N1)$)
node[midway, sloped, below, yshift=-0.75mm] {\tiny $\delta_f^3 = 10$};

\draw[->, thick, shorten >=0.15cm, shorten <=0.15cm, color=OrangeOLM] 
($(N8)!0!(N3)!3mm!25:(N3)$) -- 
($(N3)!0!(N8)!3mm!-25:(N8)$)
node[midway, sloped, below, yshift=-0.75mm] {\tiny $\delta_f^3 = 3$};

\draw[->, thick, shorten >=0.15cm, shorten <=0.15cm, color=OrangeOLM] 
($(N6)!0!(N7)!3mm!25:(N7)$) -- 
($(N7)!0!(N6)!3mm!-25:(N6)$)
node[midway, sloped, below, yshift=-0.75mm] {\tiny $\delta_f^3 = 8$};

\draw[->, thick, shorten >=0.15cm, shorten <=0.15cm, color=OrangeOLM] 
($(N7)!0!(N8)!3mm!25:(N8)$) -- 
($(N8)!0!(N7)!3mm!-25:(N7)$)
node[midway, sloped, below, yshift=-0.75mm] {\tiny $\delta_f^3 = 6$};

\end{tikzpicture}
}

\newcommand{\MultiPathDPS}[0]{
\begin{tikzpicture}[
    device/.style={rectangle, draw, minimum size=0.75cm},
    node/.style={circle, draw, minimum size=0.75cm},
    layer/.style={rectangle, draw, minimum width=0.4cm, minimum height=0.2cm},
    
    every path/.style={draw}
  ]

\node[device] (S) {\scriptsize S};
\node[node, right=1.125cm of S] (N1) {\scriptsize 1}; 
\node[node, right=3cm of N1] (N3) {\scriptsize 3}; 
\node[device, right=1.125cm of N3] (D) {\scriptsize D};
\node[node, above=0.9cm of $(N1)!0.5!(N3)$] (N2) {\scriptsize 2};
\node[node, below=1.25cm of $(N1)!0.25!(N3)$] (N4) {\scriptsize 4};
\node[node, below=1.25cm of $(N1)!0.75!(N3)$] (N5) {\scriptsize 5};

\path (S) -- (N1);
\path (N1) -- (N2);
\path (N2) -- (N3);
\path (N1) -- (N4);
\path (N4) -- (N5);
\path (N5) -- (N3);
\path (N3) -- (D);

\foreach \n in {N1, N3} {
    \node[layer, fill=BlueFillL, above=0.225cm of \n] (\n L1) {\scriptsize $l_3 = 3 $ ms};
    \node[layer, above=0cm of \n L1] (\n L2) {\scriptsize $l_2 = 2 $ ms};
    \node[layer, above=0cm of \n L2] (\n L3) {\scriptsize $l_1 = 1 $ ms};
}

\foreach \n in {N2} {
    \node[layer, fill=BlueFillL, above=0.2125cm of \n] (\n L1) {\scriptsize $l_3 = 3 $ ms};
    \node[layer, above=0cm of \n L1] (\n L2) {\scriptsize $l_2 = 2 $ ms};
    \node[layer, above=0cm of \n L2] (\n L3) {\scriptsize $l_1 = 1 $ ms};
}

\foreach \n in {N1} {
    \node[layer, below=0.225cm of \n, xshift=-0.5cm] (\n L1) {\scriptsize $l_1 = 1 $ ms};
    \node[layer, fill=GreenFillL, below=0cm of \n L1] (\n L2) {\scriptsize $l_2 = 2 $ ms};
    \node[layer, below=0cm of \n L2] (\n L3) {\scriptsize $l_3 = 3 $ ms};
}

\foreach \n in {N3} {
    \node[layer, below=0.225cm of \n, xshift=0.5cm] (\n L1) {\scriptsize $l_1 = 1 $ ms};
    \node[layer, below=0cm of \n L1] (\n L2) {\scriptsize $l_2 = 2 $ ms};
    \node[layer, fill=GreenFillL, below=0cm of \n L2] (\n L3) {\scriptsize $l_3 = 3 $ ms};
}

\foreach \n in {N4, N5} {
    \node[layer, below=0.225cm of \n] (\n L1) {\scriptsize $l_1 = 1 $ ms};
    \node[layer, fill=GreenFillL, below=0cm of \n L1] (\n L2) {\scriptsize $l_2 = 2 $ ms};
    \node[layer, below=0cm of \n L2] (\n L3) {\scriptsize $l_3 = 3 $ ms};
}

\draw[->, thick, shorten >=0.05cm, shorten <=0.15cm] ([yshift=0.15cm]S.east) -- ([yshift=0.15cm]N1.west)
node[midway, above] {\scriptsize $p = 3$};
\draw[->, thick, shorten >=0.15cm, shorten <=0.05cm] ([yshift=0.15cm]N3.east) -- ([yshift=0.15cm]D.west)
node[midway, above] {\scriptsize $p = 3$};

\draw[->, thick, shorten >=0.05cm, shorten <=0.05cm]   
($ (N1)!0!(N2) + (0.25cm+0.2cm,0.2cm-0.1cm) $) 
-- ($ (N1)!1!(N2) + (-0.25cm-0.05cm,-0.2cm-0.2cm) $)
node[midway, sloped, below] {\scriptsize $\sigma = 6$};

\draw[->, thick, shorten >=0.05cm, shorten <=0.05cm]   
($ (N2)!0!(N3) + (0.25cm+0.05cm,-0.2cm-0.2cm) $) 
-- ($ (N2)!1!(N3) + (-0.25cm-0.2cm,0.2cm-0.1cm) $)
node[midway, sloped, below] {\scriptsize $\sigma = 3$};

\draw[->, thick, shorten >=0.05cm, shorten <=0.05cm]   
($ (N1)!0!(N4) + (0.15cm+0.2cm,-0.25cm-0.1cm) $) 
-- ($ (N1)!1!(N4) + (-0.15cm+0.05cm,0.25cm+0.2cm) $)
node[midway, sloped, above] {\scriptsize $\sigma = 7$};

\draw[->, thick, shorten >=0.05cm, shorten <=0.05cm]   
([yshift=0.15cm]N4.east) -- ([yshift=0.15cm]N5.west)
node[midway, above] {\scriptsize $\sigma = 5$};

\draw[->, thick, shorten >=0.05cm, shorten <=0.05cm]   
($ (N5)!0!(N3) + (0.15cm-0.05cm,0.25cm+0.2cm) $) 
-- ($ (N5)!1!(N3) + (-0.15cm-0.2cm,-0.25cm-0.1cm) $)
node[midway, sloped, above] {\scriptsize $\sigma = 3$};

\draw[decorate, decoration={coil, aspect=0, segment length=4mm, amplitude=2mm}, ->]
        (0.9,0.9) -- (N1);

\node at (0.2,1.2) {\scriptsize $\sigma = p \cdot \hat{d}$};
\node at (0.2,0.9) {\scriptsize $ = 3 \cdot 3 = 9$};

\end{tikzpicture}
}

\section{System Model and Jitter Amplification under FRER} \label{SOTA Section}

This section defines the network and scheduling models and analyzes jitter amplification under FRER.

\subsection{Network and Flow Models} \label{Network Model}

We consider a packet-switched network composed of end hosts connected through a set of intermediate nodes and physical links. Each link, together with its associated egress port, provides a transmission capacity $C$.

Each intermediate node supports $K$ traffic classes, implemented as distinct priority queues indexed by $k \in \{1, ..., K\}$, where lower indices correspond to higher priority. Packet scheduling among traffic classes is performed using the Multi-CQF mechanism, extended with the HSSU scheduler to enable flexible cycle time configuration across classes. 

The network carries a set of flows $\mathcal{F}$. Each flow $f \in \mathcal{F}$ originates at a source node and is destined for a corresponding destination node. A priority value $P_f \in \{0,...,7\}$ is assigned to each flow, corresponding to the IEEE 802.1Q Priority Code Point (PCP). Let $\mathcal{M}: \{0,...,7\} \to \{1,...,K\}$ denote the configured mapping from PCP values to traffic classes\footnote{The PCP range is partitioned into $K$ contiguous intervals, each assigned to one traffic class in order of priority.}. The traffic class of flow $f$ is given by
\begin{equation}
    k_f = \mathcal{M}(P_f).
    \vspace{-3pt}
\label{equ:pcp to class mapping}
\end{equation}
Packets of flow $f$ are enqueued in the priority queue associated with class $k_f$ at each intermediate node along the selected forwarding path.

We focus on predictable traffic types, including periodic flows characterized by a minimum inter-arrival time and maximum burst size, and aperiodic flows, which do not follow fixed transmission intervals but generate constrained and predictable data volumes. Bursty traffic is outside the scope of this work.

\subsection{Multi-CQF Scheduling Model}

In Multi-CQF, each traffic class $k$ represents a logical queue composed of $Q_k$ cyclically served sub-queues. Throughout this paper, the term \textit{queue} refers to the traffic class indexed by $k$, whereas \textit{sub-queue} refers to the internal buffers used by Multi-CQF to realize cyclic forwarding within that class.

Each traffic class $k$ is associated with a cycle time $T_k$ and is allocated a fraction $\lambda_k$ of the outgoing link capacity $C$, subject~to
\begin{equation}
\sum_{k=1}^K \lambda_k \leq 1 .
\vspace{-3pt}
\end{equation}

Due to the cyclic forwarding structure, the worst-case queuing delay at a single hop for packets of class $k$ is bounded by:
\begin{equation}
    W_k =  Q_k \cdot T_k,
\end{equation}
and the E2E latency $L_k$ traversing a path of $h$ hops satisfies $L_{k}^{\min} \leq L_k \leq L_{k}^{\max}$, where
\begin{equation}
    L_{k}^{\min} = (h-1)\cdot T_k
    \vspace{-3pt}
\end{equation}
and 
\begin{equation}
    L_{k}^{\max} = Q_k \cdot T_k + (h-1)\cdot T_k.
    \vspace{-3pt}
    \label{equ:latencybound}
\end{equation}

The maximum E2E latency arises when the worst-case queuing delay~$W_k$ is encountered at the first hop and one cycle delay~$T_k$ is incurred at every subsequent hop. The minimum E2E latency arises when no queuing delay is experienced at the first hop and one cycle delay is incurred at each subsequent hop. Accordingly, the E2E jitter of packets belonging to class~$k$ is bounded by 
\begin{equation}
J_k = Q_k \cdot T_k.
\label{equ:jitterbound}
\end{equation}

\subsection{Jitter Amplification under FRER}\label{Jitter Amplification under FRER}

FRER replicates packets at the ingress node and forwards them over multiple disjoint paths. At the egress node, the first arriving replica is delivered while duplicates are discarded. Although FRER improves reliability, forwarding replicated packets over multiple paths with unequal delays may interfere with the timing guarantees provided by the underlying scheduling mechanism.

In Multi-CQF, a flow~$f$ is statically mapped to traffic class~$k_f$ at every hop. Hence, its E2E latency and jitter bounds are fully determined by the parameters of class~$k_f$ and the hop count $h$ of the forwarding path. These guarantees hold deterministically as long as all packets traverse a single path with fixed hop count. When FRER is enabled, replicas of the same packet may traverse routes with different hop counts and cumulative delays, while still being scheduled according to the same class $k_f$ at every hop.

Although for each individual path, the Multi-CQF jitter bound (\ref{equ:jitterbound}) still holds, heterogeneous path lengths introduce an additional latency offset between routes. Because each additional hop contributes one cycle time $T_{k_f}$, the difference between corresponding latency bounds on the shortest path (with $h=h_f^{\min}$) and the longest path ( with $h=h_f^{\max}$) for a flow $f$ equals
\begin{equation}
    (h_f^{\max} - h_f^{\min})\cdot T_{k_f}.
\end{equation}

Since the egress node delivers the first-arriving replica, packets are normally received via the shortest path. Upon failure of this path, however, subsequent packets are delivered via a longer route, resulting in a step increase in latency and increased inter-arrival variation at the egress node. Under these assumptions, the E2E delay variation bound across the redundant paths is
\begin{equation}
    J_{k_f} + (h_f^{\max} - h_f^{\min})\cdot T_{k_f}.
\end{equation}

\begin{figure}[tp]
    \centering
    
    \subfloat[Static traffic class assignment along the primary path.]{%
        \SinglePathShortestPath
        \begin{tikzpicture}[>=latex]

        \vspace{-3mm}
        \end{tikzpicture}
        \label{Subfig: Static traffic class assignment along the primary path}
    }

    \vspace{-2mm}

    \subfloat[Static traffic class assignment along redundant paths.]{%
        \MultiPathRPS
        \begin{tikzpicture}[>=latex]

        \vspace{-3mm}
        \end{tikzpicture}
        \label{Subfig: Static traffic class assignment along redundant paths}
    }

    \caption{Static traffic class assignment under Multi-CQF and FRER.}
    \label{Fig: Static traffic class assignment in Multi-CQF}
    \vspace{-3mm}
\end{figure}
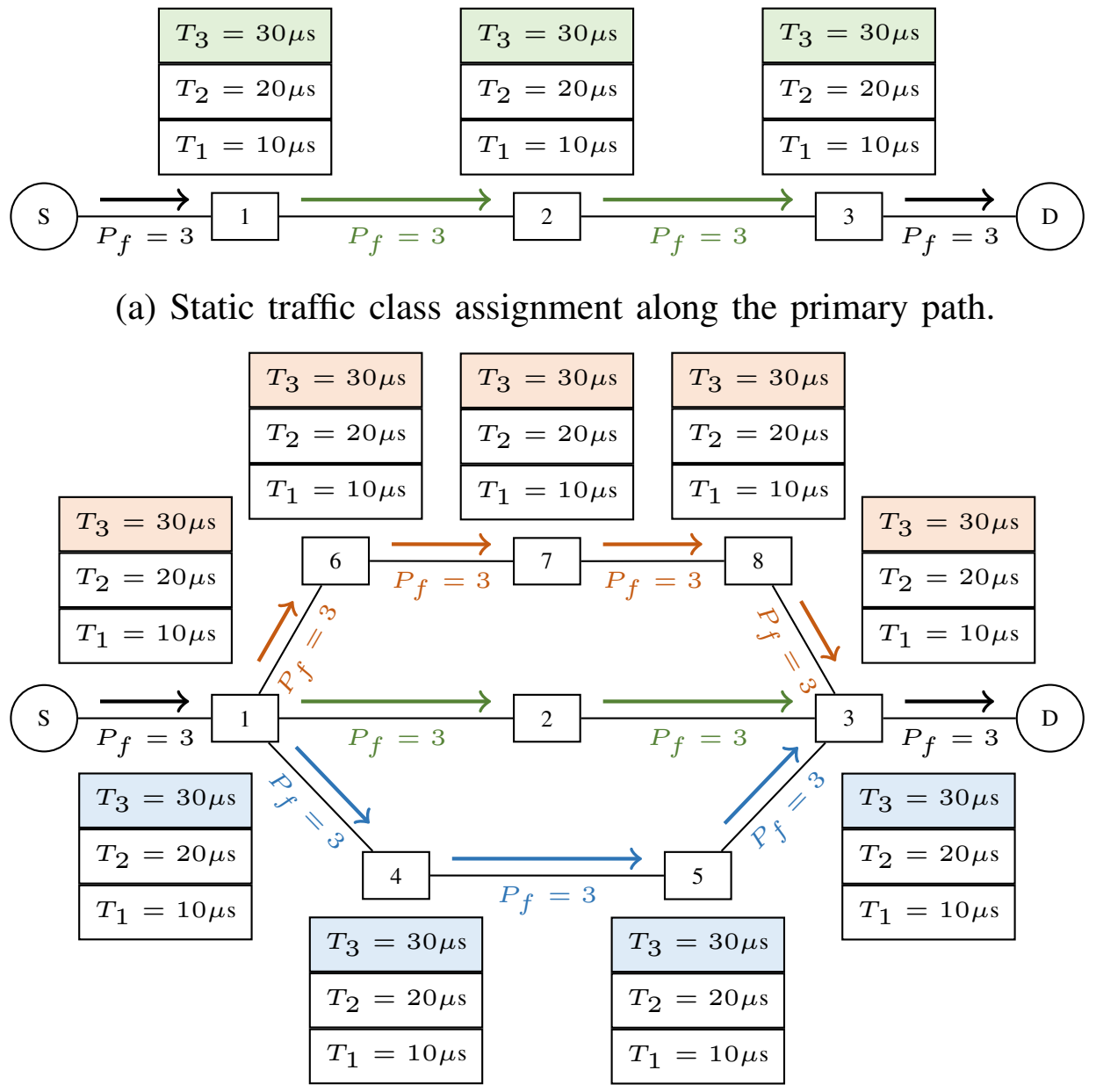

An illustrative example is shown in Fig. \ref{Fig: Static traffic class assignment in Multi-CQF}. For simplicity, we assume a one-to-one mapping between the priority value~$P_f$ and traffic class $k$, two sub-queues per class ($Q_k = 2$), and neglect source processing and propagation delays. Under single-path forwarding (Fig. \ref{Subfig: Static traffic class assignment along the primary path}), a flow $f$ with $P_f = 3$ is mapped to class $k=3$ and achieves an E2E latency within $[60,120]~\mu s$, satisfying the Multi-CQF jitter bound ($J_k = 2T_k = 60~\mu s$).

When FRER is enabled (Fig. \ref{Subfig: Static traffic class assignment along redundant paths}), packets are replicated and forwarded over three paths: $1$-$2$-$3$ (\textit{PrimaryPath}), $1$-$4$-$5$-$3$ (\textit{FirstAlternative}), and $1$-$6$-$7$-$8$-$3$ (\textit{SecondAlternative}). The corresponding E2E latency intervals shift from $[60, 120]~\mu s$ on the \textit{PrimaryPath} to $[90,150]~\mu s$ and $[120,180]~\mu s$ on the alternative routes. Although each path individually satisfies the jitter bound of~$60~\mu s$, the latency offset between paths introduces additional inter-arrival variation when failure occurs. The inter-arrival variation, between packet $i$ delivered via the \textit{PrimaryPath} with minimum latency and packet $i+1$ delivered via a redundant path with maximum latency, increases to $90~\mu s$ (\textit{FirstAlternative}) or $120~\mu s$ (\textit{SecondAlternative}), exceeding the guaranteed jitter bound of $60~\mu s$.

\section{Quota-Based Traffic Classification for Multi-CQF under FRER} \label{Novel Section}

This section presents a quota-based dynamic traffic classification mechanism for Multi-CQF under FRER. Each replica is assigned a common latency budget derived from static Multi-CQF on the primary path. Traffic classes are selected hop-by-hop according to the remaining budget and hop count, thereby mitigating latency differences and jitter amplification across redundant paths.

The following subsections describe quota initialization, traffic classification at the ingress and subsequent nodes, and duplicate elimination at the egress node.

\subsection{Quota Initialization at Ingress Node}
\label{sec:Quota Initialization at the Ingress Node}

For each flow $f \in \mathcal{F}$, let $\mathcal{R}_f = \{r_1,\dots,r_{R_f}\}$ denote the set of candidate routes to the destination, where $R_f = |\mathcal{R}_f|$. At the ingress node, flow $f$ is replicated into $R_f$ replicas, each forwarded along a distinct route $r \in \mathcal{R}_f$.

Let $h_f^r$ denote the hop count of flow $f$ on route $r$, and~$h_f^{\min}$ be the minimum hop count across all routes, i.e., the hop count on the shortest path. Nodes along route $r$ are indexed sequentially by $n \in \{1,\dots,h_f^r\}$, where $n=1$ corresponds to the first hop and $n=h_f^r$ to the last hop.

Using the static priority field, $P_f$, assigned to flow $f$, the ingress node determines the reference traffic class
\begin{equation}
    \hat{k}_f = \mathcal{M}(P_f),
    \vspace{-3pt}
\end{equation}
according to the predefined priority-to-class mapping in~(\ref{equ:pcp to class mapping}). This reference class corresponds to the traffic class that would be used under static Multi-CQF scheduling.

Two quotas, normalized by the highest-priority cycle time $T_1,$ are initialized from the reference class $\hat{k}_f$ and shortest-path hop count $h_f^{\min}$. The first quota, $\Delta_f^r$, is used for ingress class selection and assumes one reference cycle per hop:
\begin{equation}
    \Delta_f^r = \frac{T_{\hat{k}_f}}{T_1}\cdot h_f^{\min} ,
    \vspace{-3pt}
\end{equation}
where $T_{\hat{k}_f}$ is the cycle time of the reference traffic class $\hat{k}_f$ and~$T_1$ is the cycle time of the highest-priority class. 

The second quota, $\delta_f^r$, additionally accounts for the first-hop contribution of up to $Q_{\hat{k}_f}$ reference cycles:
\begin{equation}
    \delta_f^r = Q_{\hat{k}_f} \cdot \frac{T_{\hat{k}_f}}{T_1} + (h_f^{\min}-1)\cdot \frac{T_{\hat{k}_f}}{T_1}.
\end{equation}

Crucially, both quotas are computed from the same $h_f^{\min}$ and $\hat{k}_f$, establishing a common E2E latency budget across all replicas. The quota $\Delta_f^r$ governs ingress classification, while $\delta_f^r$ is updated independently along each route to guide classification at subsequent hops.

\subsection{Traffic Classification at Ingress Node}
\label{sec:Traffic Classification at the Ingress Node}

To reflect the dynamic per-replica, per-hop assignment performed by the proposed mechanism, we denote by $k^{(r,n)}$ the traffic class assigned to the replica traversing route $r$ at node~$n$.

At the ingress node, two steps are performed. First, the initial traffic class of each replica is determined using $\Delta_f^r$, assuming a uniform distribution of the prioritization budget over the corresponding route. Second, after class selection, the  quota~$\delta_f^r$, which regulates traffic classification at subsequent hops, is updated to account for the worst-case latency contribution incurred at the first hop under Multi-CQF, i.e.,~$Q_{k^{(r,n)}} T_{k^{(r,n)}}$.

To determine the traffic classification at the ingress node ($n=1$), the maximum admissible cycle time for a replica routed over~$r$ with $h_f^r$ hops is given by

\begin{equation}
    \phi_f^{(r,~n=1)} = \frac{\Delta_f^r}{h_f^r} \cdot T_1.
\end{equation}

Starting from the lowest-priority class, the first class whose cycle time satisfies the following condition is selected:
\begin{equation}
    T_{k^{(r,n)}} \leq \phi_f^{(r,n)}.
\label{equ:class selection}
\end{equation}
Because the hop count $h_f^r$ may differ across routes, replicas of the same flow can be assigned different traffic classes at the ingress node.

After selection, $\Delta_f^r$ is discarded and $\delta_f^r$ is reduced by the normalized first-hop latency contribution:
\begin{equation}
    \delta_f^r \to \delta_f^r - Q_{k^{(r,1)}} \cdot \frac{T_{k^{(r,~n=1)}}}{T_1}.
    \vspace{-3pt}
\label{equ: residual update at ingress node}
\end{equation}
The updated quota $\delta_f^r$ is then embedded into each packet, extending the IEEE 802.1Q tag, and carried to subsequent hops for dynamic hop-by-hop classification.

\subsection{Hop-by-Hop Traffic Classification at Subsequent Nodes}
\label{sec:Hop-by-Hop Traffic Classification at Subsequent Nodes}

At each subsequent node $n$, where $n > 1$, the residual quota $\delta_f^r$ is distributed over the remaining $h_f^r-(n-1)$ hops to determine the maximum admissible per-hop latency contribution:
\begin{equation}
    \phi_f^{(r,~n>1)} = \frac{\delta_f^r}{h_f^r-(n-1)} \cdot T_1.
\label{equ:allowable subsequent latency}
\end{equation}
Similar to the procedure at the ingress node, the lowest-priority class satisfying~(\ref{equ:class selection}) is selected.

After classification, the residual quota is updated as
\begin{equation}
    \delta_f^r \xleftarrow{} \delta_f^r - \frac{T_{k^{(r,n)}}}{T_1},
\label{equ:deduction}
\end{equation}
accounting for the worst-case latency contribution at that hop. Unlike the first-hop update in~(\ref{equ: residual update at ingress node}), subsequent updates deduct only one cycle time.

\subsection{Classification and Duplicate Elimination at Egress Node}
\label{sec:Egress Node}

At the egress node, duplicate elimination retains only the first-arriving replica of each frame. This replica is classified using (\ref{equ:allowable subsequent latency}) and (\ref{equ:class selection}), as at other intermediate nodes. The quota field is then removed, restoring the packet to its original Ethernet frame format. Notably, if the deduction in (\ref{equ:deduction}) were applied, the residual quota would ideally be fully consumed, confirming equivalent worst-case E2E latency across the redundant paths.

\begin{figure}[tp]
    \centering
    \MultiPathQuota
    \begin{tikzpicture}[>=latex]

    \end{tikzpicture}
    \vspace{-2mm}
    \caption{Quota-based traffic classification and hop-by-hop quota updates in Multi-CQF under FRER.}
    \label{Fig:Quota-based traffic classification in Multi-CQF under FRER}
    \vspace{-3mm}
\end{figure}
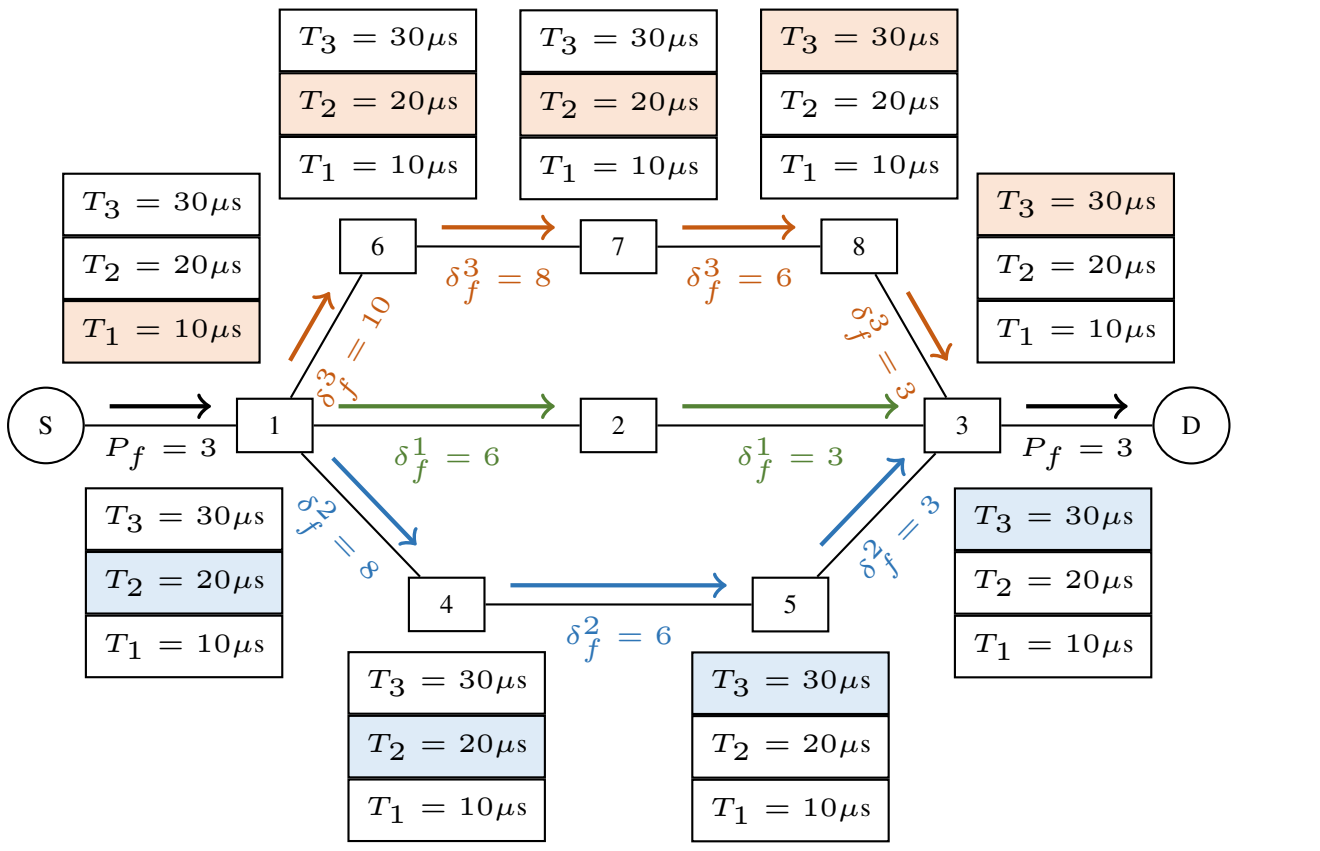

Fig. \ref{Fig:Quota-based traffic classification in Multi-CQF under FRER} illustrates the proposed mechanism using the same example introduced in Section~\ref{Jitter Amplification under FRER}. At the ingress node, the quotas~$\Delta_f^r = 9$ and $\delta_f^r = 12$ are initialized and attached to all replicas of flow $f$ routed over the \textit{PrimaryPath}, \textit{FirstAlternative}, and \textit{SecondAlternative}.

On the \textit{PrimaryPath}, traffic classification remains static with $k^{(1,n)} = 3$ at all hops. In contrast, replicas on the redundant paths are dynamically classified according to the residual quota: $k^{(2,n)} = [2,2,3,3]$ on the \textit{FirstAlternative} and~$k^{(3,n)} = [1,2,2,3,3]$ on the \textit{SecondAlternative}.

The resulting E2E latency intervals are $[60, 120]~\mu s$ over the \textit{PrimaryPath}, $[80, 120]~\mu s$  over the \textit{FirstAlternative}, and $[100, 120]~\mu s$ over the \textit{SecondAlternative}. Unlike static classification, the maximum E2E latency is identical across all routes. Consequently, under failure, the inter-arrival variation remains bounded by $60~\mu s$, satisfying the Multi-CQF jitter guarantee.

\section{Evaluation} \label{Evaluation}

This section evaluates the performance of the proposed quota-based traffic classification mechanism against single-path Multi-CQF and Multi-CQF with FRER under static classification. The evaluation is conducted using OMNEST~\cite{omnest}.

\subsection{Simulation Setup}

We consider the network topology shown in Fig.~\ref{Fig:Quota-based traffic classification in Multi-CQF under FRER}, the same topology used in the illustrative example. The network consists of multiple sources connected to an ingress node and a single destination connected to an egress node. The network provides redundant paths to the destination, where the shortest path includes three intermediate nodes. All links operate at 1~Gbps. Intermediate nodes employ Multi-CQF with the HSSU scheduler, supporting four traffic classes ($K = 4$), each with four sub-queues ($Q_k = 4$) and per-sub-queue buffer capacity~$\sigma_k$. Configuration parameters are shown in Table~\ref{tab:Simulation Configuration Parameters}.

The simulation includes heterogeneous traffic types, following the IEC/IEEE 60802 TSN Profile for Industrial Automation \cite{IEC60802}. Class~1 carries time-triggered traffic with periods ranging from $50$--$200~\mu$s. Classes 2 and 3 carry periodic traffic with intervals of $250$--$500~\mu$s and packet sizes of $500$--$1000$ bytes, as well as aperiodic traffic with exponential inter-arrival times and fixed packet sizes of $500$--$1500$ bytes. Class 4 carries background traffic.

Time-triggered traffic is strictly forwarded over the primary path. For classes 2--4, we evaluate three configurations: (i)~Multi-CQF on a single path, (ii) Multi-CQF with FRER under static classification, and (iii) Multi-CQF with FRER under the proposed quota-based classification.

\begin{table}[!t]
\centering
\renewcommand{\arraystretch}{0.5}
\setlength{\tabcolsep}{9pt}

\caption{Simulation Configuration Parameters.}
\vspace{-1mm}
\label{tab:Simulation Configuration Parameters}
\begin{tabular}{c|cc|cc}
\toprule
\textbf{Class} & \multicolumn{2}{c|}{\textbf{Configuration}} & \multicolumn{2}{c}{\textbf{Bounds}} \\ 

 \( k \) & \(T_k\) & \(4\cdot\sigma_k\) & \(J_k\) & \(L_k^{\max}\) \\
 
\midrule

1 & \(\ 100~\mu\text{s}\) & \( 100\ \text{kb}\) & \(\ 400~\mu\text{s}\) & \(\ 600~\mu\text{s}\)\\ 

2 &\(\ 200~\mu\text{s}\) & \( 200\ \text{kb}\) & \(\ 800~\mu\text{s}\) & \(\ 1200~\mu\text{s}\)\\ 

3 & \(\ 300~\mu\text{s}\) & \( 300\ \text{kb}\) & \(\ 1200~\mu\text{s}\) & \(\ 1800~\mu\text{s}\)\\ 

4 & \(\ 400~\mu\text{s}\) & \( 400\ \text{kb}\) & \(\ 1600~\mu\text{s}\) & \(\ 2400~\mu\text{s}\)\\ 

\bottomrule
\end{tabular}
\vspace{-4mm}
\end{table}

\subsection{Simulation Results}
\begin{figure}[tp]
    \centering
    \includegraphics[width=0.9\linewidth]{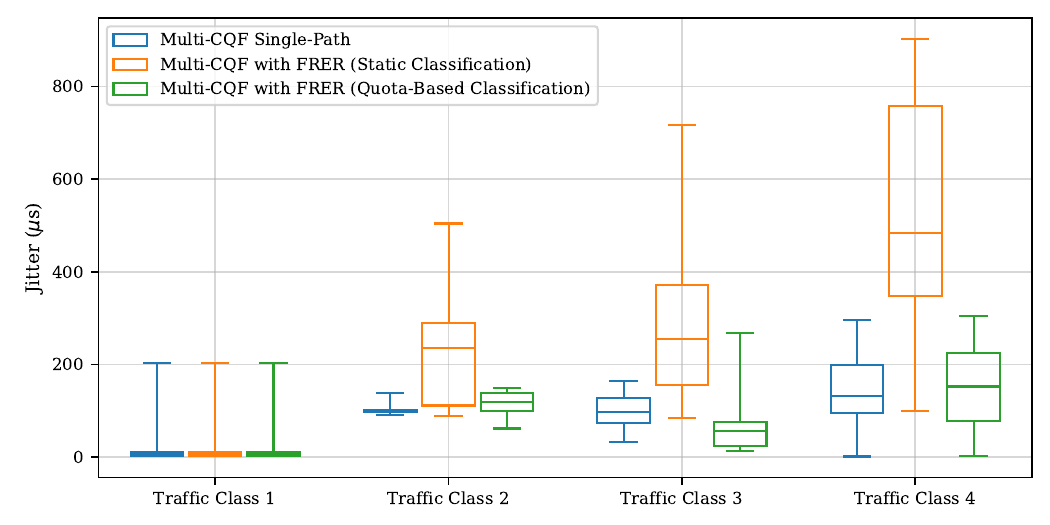}
    \vspace{-3mm}
    \caption{E2E jitter per traffic class under single-path forwarding, static FRER classification, and quota-based FRER classification.}
    \label{fig:jitter}
    \vspace{-3mm}
\end{figure}

\begin{figure}[tp]
    \centering
    \includegraphics[width=0.9\linewidth]{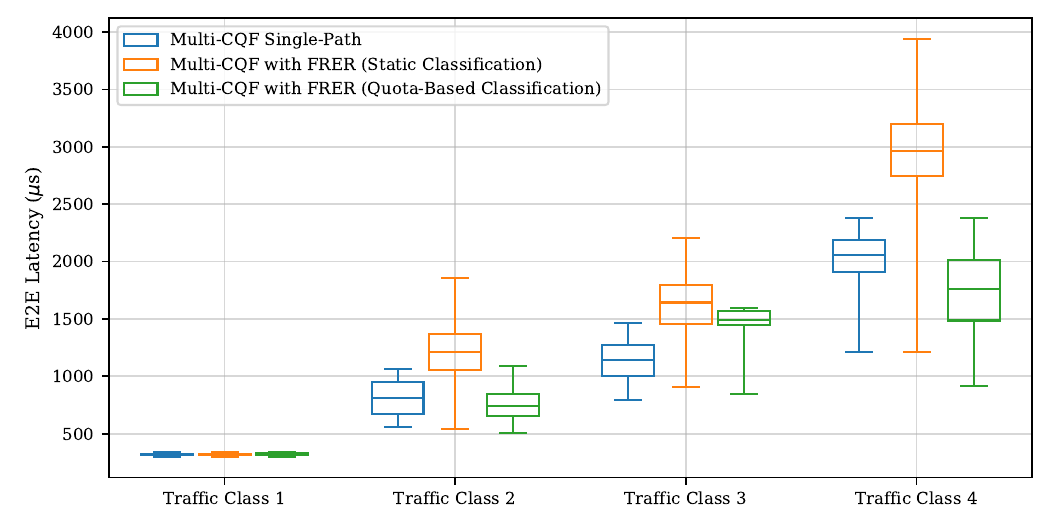}
    \vspace{-3mm}
    \caption{E2E latency per traffic class under single-path forwarding, static FRER classification, and quota-based FRER classification.}
    \label{fig:Latency}
    \vspace{-3mm}
\end{figure}

Figs.~\ref{fig:jitter} and \ref{fig:Latency} present the observed E2E jitter and latency per traffic class. The whiskers denote the minimum and maximum values, the box represents the interquartile range (25th--75th percentiles), and the horizontal line indicates the mean. 

When FRER is enabled with static traffic classification, both E2E jitter and latency increase significantly compared to single-path forwarding. The degradation is most pronounced on the longest redundant path. Across classes 2--4, jitter increases by up to 337\%, while latency increases by up to 74\%. This asymmetry is because the jitter bound~(\ref{equ:jitterbound}) is comparatively small relative to the maximum latency bound~(\ref{equ:latencybound}).

In contrast, when the proposed quota-based classification is applied, the performance degradation under FRER is markedly reduced. The maximum jitter increase relative to single-path forwarding is limited to 63\%, and for most classes remains below 10\%. Similarly, latency increases by at most 9\% compared to single-path forwarding. Relative to static classification, the proposed mechanism reduces FRER-induced jitter amplification by up to 70\% and latency degradation by up to 41\%. 

\section{Conclusion} \label{Conclusion}

This paper investigated the jitter amplification problem arising when combining Multi-CQF with FRER over redundant paths with unequal hop counts. To address this, we proposed a quota-based traffic classification mechanism that adapts class assignments hop-by-hop using a latency budget derived from the primary path. Simulation results confirmed that the proposed mechanism reduced jitter amplification by up to 70\% and latency degradation by up to 41\% compared to static classification. Future work will examine larger topologies, varying traffic loads, and implementation overhead.

\bibliographystyle{IEEEtran}
\bibliography{IEEEabrv,reference}

\end{document}